\documentclass[preprint,proceedings]{rmaa}

\suppressfulladdresses 

\usepackage{paralist}

\usepackage{psfrag,color}

\SetYear{2006}
\SetConfTitle{CSM and Late Stages of Massive Stellar Evolution}

\title{Cas\,A: The Bright X-ray Knots and Oxygen Emission} 

\author{
  D. Dewey,\altaffilmark{1} 
  T. DeLaney,\altaffilmark{1} 
  and J.S. Lazendic\altaffilmark{2}  }

\altaffiltext{1}{MIT Kavli Institute,
  MIT Room NE80-6085, Cambridge,
  Massachusetts 02139, USA (dd@space.mit.edu).}
\altaffiltext{2}{School of Physics,
University of Melbourne,
Victoria 3010, Australia.}

\shortauthor{Dewey, DeLaney, \& Lazendic}
\shorttitle{RevMexAA(SC) CSM and Late Stages...}

\listofauthors{D. Dewey, \& T. DeLaney}
\indexauthor{Dewey, D.}
\indexauthor{DeLaney, T.}
\indexauthor{Lazendic, J.S.}

\abstract{\textit{Chandra} High-Energy Transmission Grating (HETG)
X-ray spectra are extracted from 17 bright, narrow
regions of Cas\,A and provide unique measurements of their kinematic
and plasma states.  From the
dominant emission lines, e.g. He-like Si, we derive accurate
Doppler shifts in the range $-$2500 to $+$4000~km\,s$^{-1}$;
these agree well
with transverse-velocity measurements and allow
the features to be located in 3D.  Plasma diagnostics of these
regions indicate temperatures largely around 1\,keV with some
above 3\,keV.  Using as well the non-dispersed zeroth-order data,
we determine NEI model parameters for the regions which lead to
density estimates.  Values of $n_e\approx 100\,\mathrm{cm}^{-3}$
are likely the maximum of a range of densities in this X-ray emitting  material.
The common ``oxygen-rich'' assumption is coarsely tested by comparing
the integrated \ion{O}{8} line flux and continuum levels.  It appears
that most of the continuum is due to another source, e.g., from He.}

\addkeyword{Radiation mechanisms: Thermal}
\addkeyword{Supernova remnants: Individual(Cas~A (G111.7--2.1))}
\addkeyword{Techniques: Spectroscopic}
\addkeyword{X-rays: ISM}

\begin{document}
\maketitle

\section{HETG Observation of Cas\,A}
\label{sec:HETGobs}

The X-ray emitting material in young supernova remnants is the
bulk of the ejected mass and gives direct information on the
progenitor, the explosion mechanism, and illuminates the overall
hydrodynamics at the present epoch.  Cas\,A is a bright, Galactic
remnant which allows detailed study in all wavebands.  Here we
add high spatial and spectral resolution
X-ray measurements to the Cas\,A knowledge base.

\begin{figure*}[!t]\centering
  %
  \includegraphics[width=\textwidth]{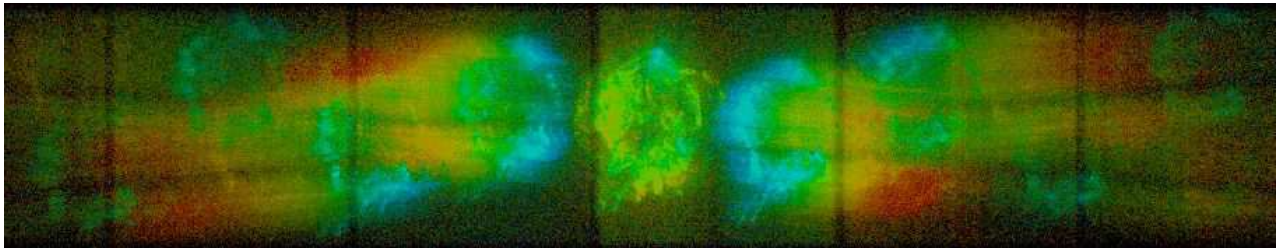}
  \caption{Dispersed image of Cas\,A on the ACIS-S array.  The undispersed
``zeroth-order'' image of Cas\,A is seen near the center; with axes roughly
N to left and E down.  Dispersed images of Cas\,A  in the bright Si emission
are visible to either side of the central image.
The MEG spectrum extends from upper-left to
lower-right and the HEG from lower-left to upper-right.
Five vertical low-intensity regions indicate the gaps between
the six ACIS CCD chips. 
The red-to-blue color range corresponds to 0.5\,keV to 2.3\,keV.}
  %
  %
  \label{fig:casadisp}
\end{figure*}

The HETG is a slitless, dispersive spectrometer~\citep{hetg05}---nominally
designed for point source observations. In contrast,
Cas\,A is a very extended source compared to the HETG 
dispersion scale, Figure~\ref{fig:casadisp}; the spacing
between the bright Si  dispersed images is not much larger than
Cas\,A itself.
Note that streaks of intensity
variation are seen along the dispersed images, the most prominent low-intensity
streaks form a wide ``X'' pattern\footnote{These streaks were brought to our attention
by Peter Ford of the ACIS science team---he initially feared that they may
be instrumental in origin!} and are likely due
to a region of high local absorption (indicated on Fig.~\ref{fig:oximage}.)

\begin{figure}[bht]\centering
  \includegraphics[bb=25 48 640 570,clip,width=0.98\columnwidth]{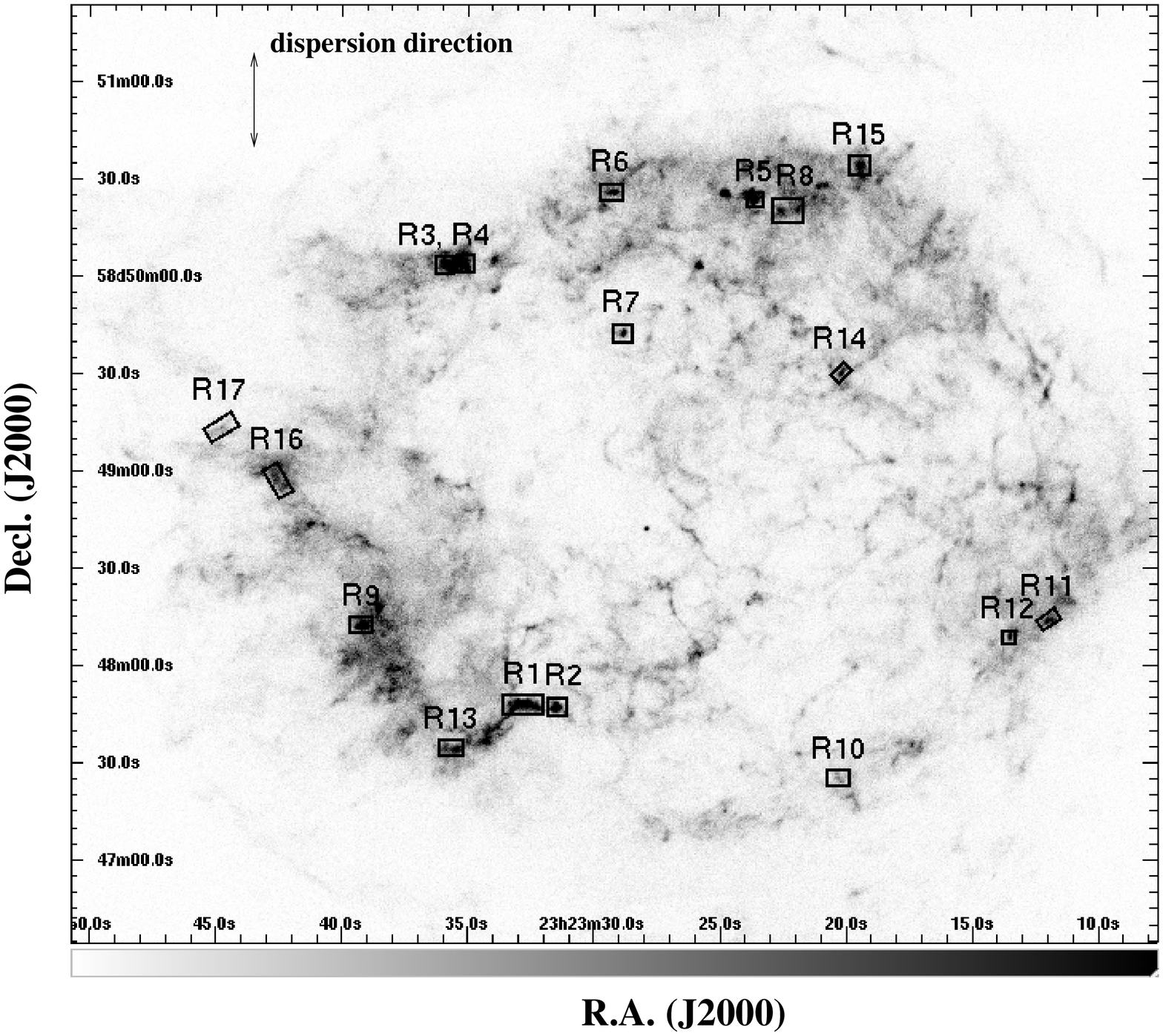}
  \caption{The 17 regions are indicated on an X-ray image of Cas\,A
(J2000 coordinates), from \citet{Lazendic06}.}
  \label{fig:Lfig2}
\end{figure}

In spite of Cas\,A's large size, we can obtain information
from these dispersed data using general techniques as described in \citet{Dewey02}.
In particular, detailed analyses were carried out for Cas\,A's bright Si line
emission and are presented in \citet{Lazendic06},
hereafter Paper\,I.
In this present work we summarize those results
in \S\,\ref{sec:velocities}\ \&\ \S\,\ref{sec:plasma},
and discuss further aspects in \S\,\ref{sec:density}\ \&\ \S\,\ref{sec:oxygen}.
Finally, future plans in these areas are discussed in \S\,\ref{sec:future}.

\section{Knot Velocities}
\label{sec:velocities}

To the HETG, Cas\,A appears as a multitude of sources; for analysis,
we selected 17 small, bright knots (here, ``knots'' is used interchangeably
with other terms: regions, features, filaments...), as indicated on
Figure~\ref{fig:Lfig2}.
We used the filament-analysis technique (Paper\,I) to narrow the features
and increase the spectral resolution.  The custom-created
PHA and RMF files were then fit in ISIS~\citep{Houck02}
to obtain line wavelengths and flux ratios.

\begin{figure}[!b]\centering
  \includegraphics[width=0.55\columnwidth]{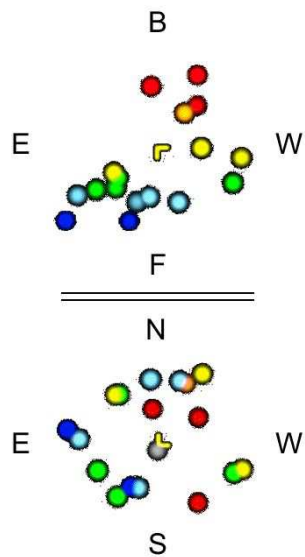}
  \caption{The regions in 3D. Bottom: cartoon version of the regions
on the sky.
Top: the same regions viewed from ``above'' with the axes E--W
and Front--Back indicated.}
  \label{fig:regs3d}
\end{figure}

Analysis of the line shifts of the bright He-like Si lines
gives accurate Doppler velocities,
in agreement with the transverse proper motions of \citet{DeLaney04}.
Assuming a uniform-expansion velocity field, the regions can
be assigned a line-of-sight z-coordinate:
$z = {\mathrm{const}} \times V_{\mathrm{meas}}$,
and the regions can be located in 3D,
as shown in Figure~\ref{fig:regs3d}.
The trend to blue-shifted velocities in the SE agrees qualitatively with
early dispersive spectroscopy results \citep{Markert83}.

\section{Plasma Properties}
\label{sec:plasma}

\begin{figure}[htb]\centering
  \includegraphics[width=\columnwidth]{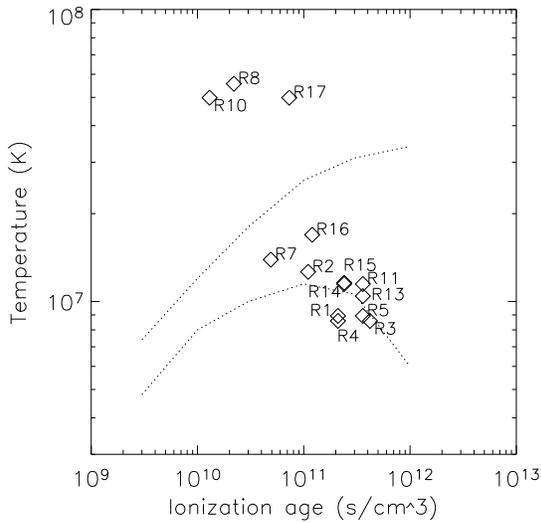}
  \caption{Regions located in T vs.~$\tau$ space;
dotted lines indicate the min--max range of the model
expectations shown in \citet{Laming03}.}
  \label{fig:laming}
\end{figure}

The $f/r$ and H-like/He-like line ratios were used with
an NEI-model-generated line-ratio grid to determine values of kT and
ionization age, $\tau = n_e t$, for the regions studied.
These values can be displayed on a ``Laming plot'',
Figure~\ref{fig:laming}.
Note that unlike the knot series measured in \citet{Laming03},
the knots here were selected for their brightness and
small size and so we don't expect them to trace out the full
extent of a model curve but rather they should be located at
the places of maximum Si-line emission along the model curves.
The three high-temperature regions,
R8, R10, and R17, clearly stand out from the general
model expectations; R17 appears to be involved with the NE jet
phenomenon, while R8 and R10 do not at first glance show a
reason for their high temperatures.

Fixing the kT and $\tau$ values, we then fit the zeroth-order
spectrum of each region 
in the range above 1.1\,keV to determines the
model ``norm'' and abundance values.
The continuum in the model fitting was assumed
to come from oxygen \citep{Laming03}, see however the discussion
in \S\,\ref{sec:oxygen}.
Using a distance to Cas\,A of 3.4\,kpc and estimated volumes for the regions,
the NEI model values can be converted to densities of the 
ions and electrons in the regions (Paper\,I, Appendix B.)

\begin{figure}[htb]\centering
  \includegraphics[width=\columnwidth]{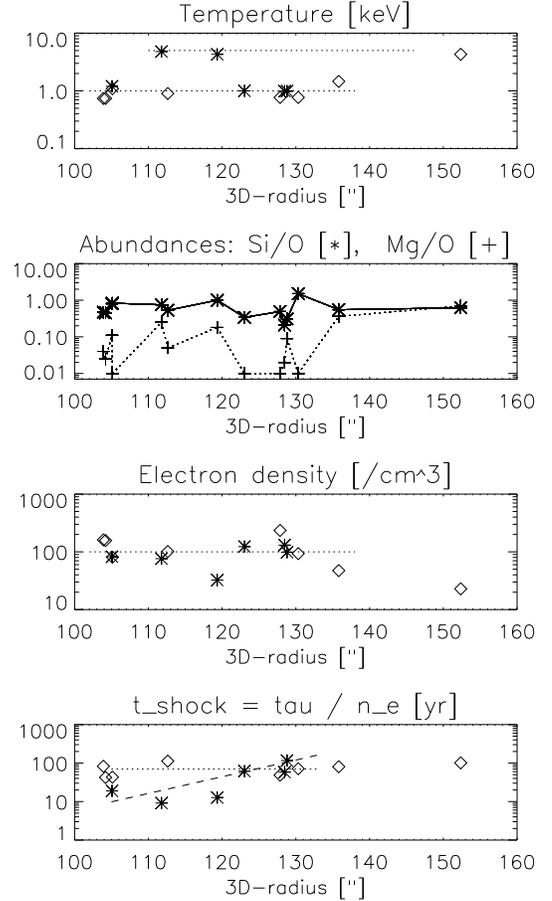}
  \caption{Properties of the regions plotted vs.~$r_{\mathrm{3D}}$.
Unless otherwise indicated, the ``*'' and diamond symbols correspond
to regions on the back and front of the remnant, respectively.
Lines on the plots serve to ``guide the eye.''
(For clarity, the outlying region R17 is plotted at a radius of 152\arcsec\ 
instead of its actual value of 172\arcsec.)}
  \label{fig:propsr3d}
\end{figure}

The derived properties of these regions are plotted as a function
of their 3D radius from the expansion center, Figure~\ref{fig:propsr3d}.
There does not appear to be any clear trend of temperature or density
with this radius as simple models (next section) might suggest.  The two outer-most
regions, R16 and R17, show solar-like abundances of Mg whereas the inner
regions have Mg at generally sub-solar levels.  In the plot of
time-since-shocked, $t_{\mathrm{shock}} = \tau / n_e$, the backside
regions closer to the center show lower values suggestive of
their having more recently encountered the reverse shock (RS).  

\section{Ejecta Density Distribution}
\label{sec:density}

The ejecta densities of the measured regions are
of order 100~cm$^{-3}$; assuming this level is typical
throughout the full
3D shell from 100\arcsec\ to 130\arcsec\ radius
leads to a very large mass estimate of the continuum-providing
oxygen (Paper\,I.)
How do these inferred densities compare with expected densities and
the generally assumed model picture?  

\begin{figure}[htb]\centering
  \hspace{0.19\columnwidth}~
  \includegraphics[width=0.4\columnwidth]{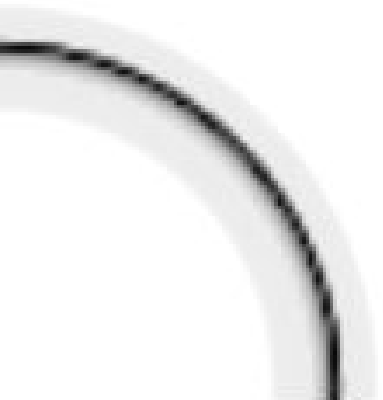}~
  \includegraphics[width=0.4\columnwidth]{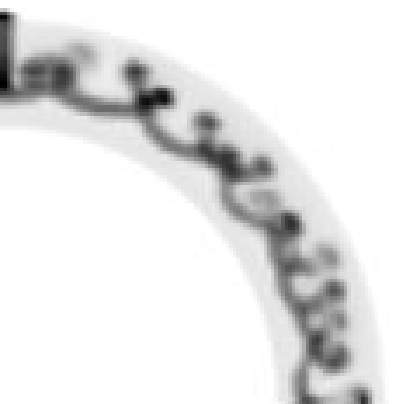}\\
  \includegraphics[width=\columnwidth]{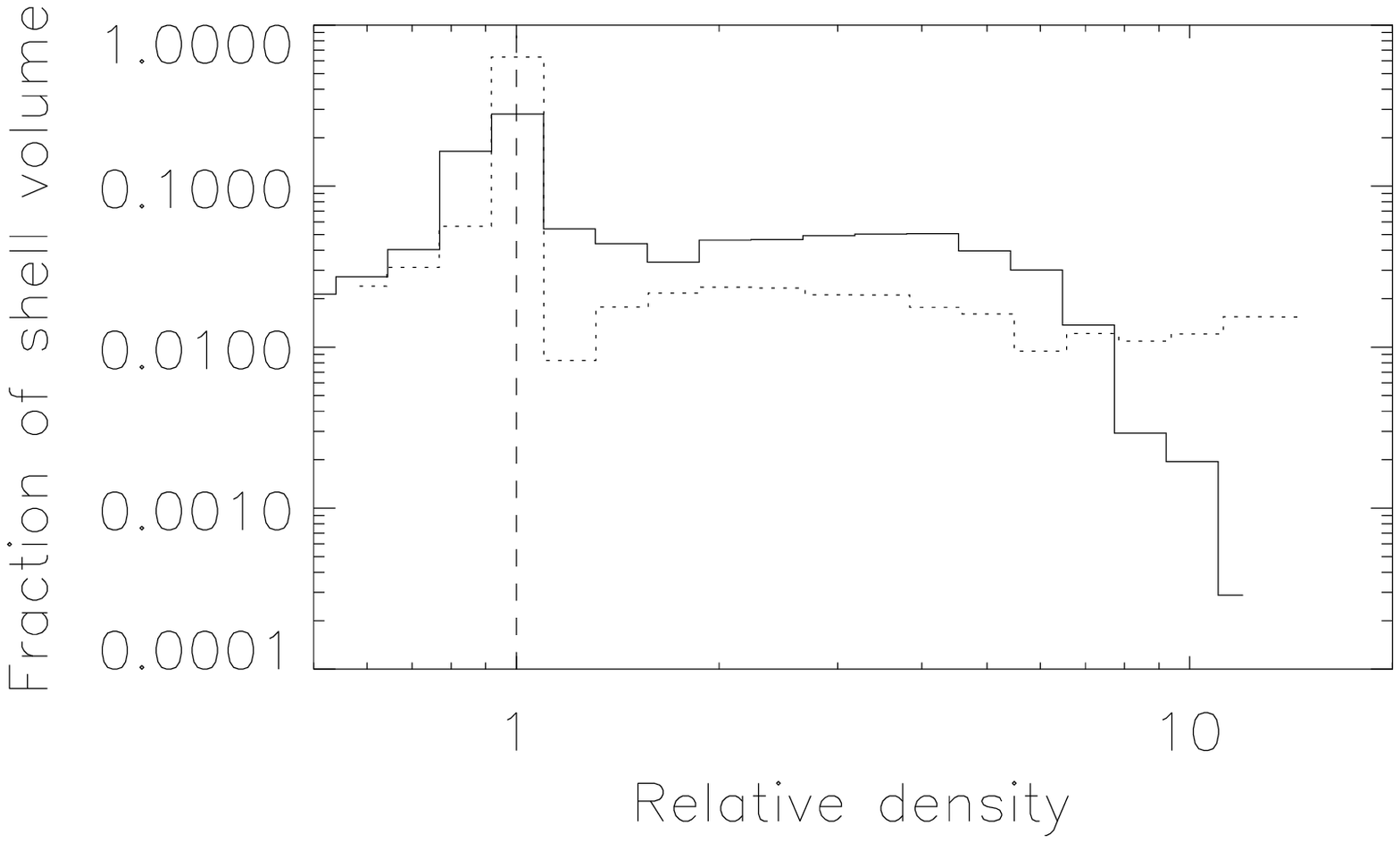}
  \caption{Histograms of model ejecta densities.
The images at upper left and right show cross-sections of a simple
self-similar distribution and an R-T-hydro density distribution, respectively.
Histograms
of the densities in a 3D shell from these models are plotted below;
the R-T-hydro simulation is the solid line.  Both models show a range of densities
extending to 10 or more times the most common value.}
  \label{fig:rtdist}
\end{figure}

The high temperature X-ray gas should trace out the bulk hydrodynamics of
the remnant, as opposed to the generally un-coupled optical FMKs and
the more mysteriously coupled (magnetic fields and relativistic electrons) radio emission.
The self-similar solution, e.g., \citet{ChevOishi},
for an $s=2$ ``wind'' circumstellar medium shows a range of densities
in the shocked ejecta with highest density at the
contact discontinuity (CD, see their Figure 2.) 
Such a trend of density with radius is not seen in our data.  However,
as suggested by Laming (private communication),
the radial location of the RS and CD
may vary around the remnant and so the effective
``thickness'' of the high-density shell is much smaller than simply
the range of radii over which we see high densities.
Further, it is likely that Rayleigh-Taylor (R-T) instabilities
will distort the otherwise
sharp CD~\citep{Chevalier92,Dwarkadas00}; note that even with R-T effects
a similarly large range of ejecta densities
may still be maintained, Figure~\ref{fig:rtdist}.
The bright, dense regions measured here then reflect the higher-density
portions of the hydro structures developed by the ejecta between the RS and CD.
Extrapolating from their density values to an overall mass estimate requires
further assumptions of the underlying density distribution.

\section{Oxygen-rich ?}
\label{sec:oxygen}

\begin{figure}[!t]\centering
  \includegraphics[width=0.8\columnwidth]{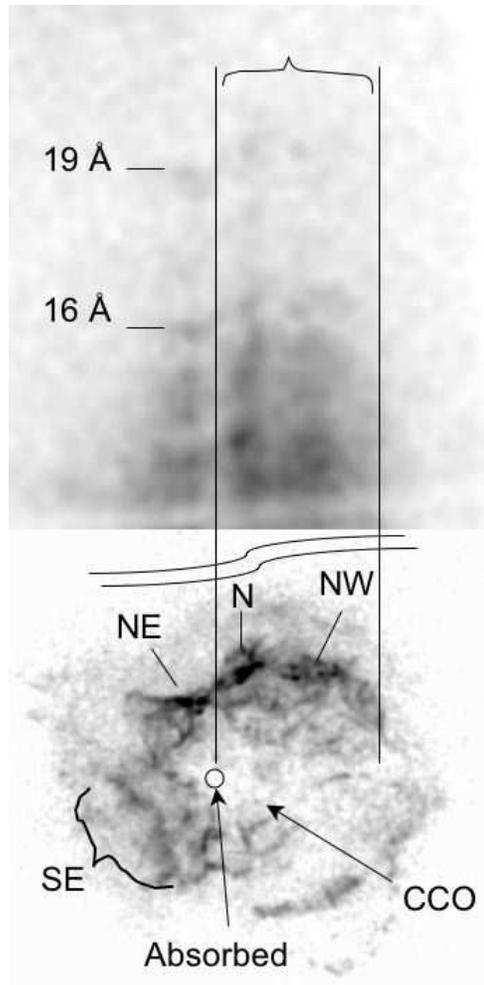}
  \caption{Oxygen in Cas\,A.  Bottom image: the oxygen band, 0.5\,keV to
0.8\,keV, from the non-grating observation, obsid 5196.  The main
O-bright regions are labeled.
Upper image: the MEG-dispersed events on S1 (binned and smoothed)
from obsid 1046; discrete \ion{O}{8} line emission is just visible from
portions of the NE, N, and NW regions.  Vertical lines indicate the extraction
region for the spectrum of Figure~\ref{fig:oxspect}.  }
  \label{fig:oximage}
\end{figure}

It has been noted that bright ejecta-dominated regions
of Cas\,A cannot be modeled
with a pure high-Z metals plasma due to a need for additional continuum
flux from elements with ``$Z<8$''~\citep{Hughes00}.
The continuum could arise from mixed-in outerlayer material:
He, N, perhaps C~\citep{PerezRendon02}.
In \citet{Laming03} and our Paper\,I this continuum
was assumed to come from oxygen for modeling purposes.

One reason that the actual amount of oxygen is an unsettled issue is that
Cas\,A is strongly absorbed with an $N_\mathrm{H}$ of
order of $10^{22}$\,cm$^{-2}$, greatly reducing the \ion{O}{8} emission.
Even so, the ACIS imaging data does include
oxygen-band emission which shows greatest
intensity in four general regions: labeled NE, N, NW and SE
in the bottom image of Figure~\ref{fig:oximage}.

It is possible to look for and
set limits on the oxygen emission using the \ion{O}{8}
Ly\,$\alpha$ and Ly\,$\beta$ lines seen in the MEG spectrum of Cas\,A;
these lines were also measured with the XMM-Newton RGS~\citep{Bleeker01}.
In that work the L$\alpha$-to-L$\beta$ ratio was measured versus cross-dispersion
location, separating the SE, NE, N regions.
We defer a similar detailed
analysis and instead consider here the
global spectrum from the combined N\,\&\,NW region.

\begin{figure}[!t]\centering
  \includegraphics[width=\columnwidth]{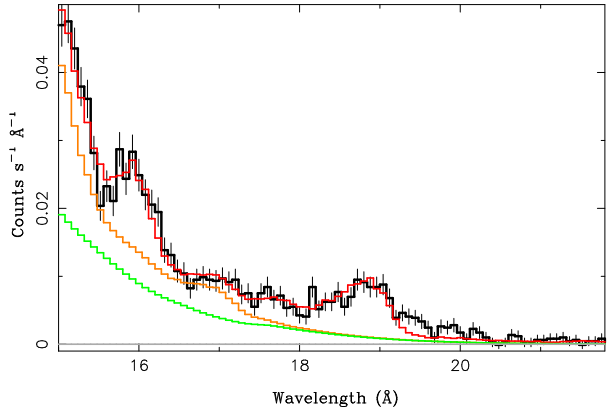}
  \caption{Oxygen seen with the MEG.  The histogram extracted
from the extended N\,\&\,NW region is well-fit by a set of XSPEC \texttt{vnei}
models which include He, Fe, and O.  The underlying He and He+Fe components
are also shown; the model is for the $kT=0.7$\,keV
and $\tau =0.2\times 10^{11}$\,s\,cm$^{-3}$
case of Table~\ref{tab:fits}. }
  \label{fig:oxspect}
\end{figure}

The extracted counts
spectrum using our filament technique is shown in 
Figure~\ref{fig:oxspect}.
This spectrum is fit by an absorbed \texttt{vnei}\ model
containing only He, Fe, and O.  The observed lines
are broader than the spatial image and velocity
components of $+1500$\,km\,s$^{-1}$ and 
$-2100$\,km\,s$^{-1}$ are included.
The difficulty of working
in this heavily absorbed range with relatively low resolution
data is demonstrated by the fitting results 
given in Table~\ref{tab:fits}.
The table shows that comparably good fits can be obtained over a large
(but relevant) range of $kT$ and $\tau$, hence, the data
really constrain only four of the six model parameters.
Note that $N_\mathrm{H}$ is considered a free parameter here
both for illustration of its effect on the fit parameters and
to allow the possibility that it differs from the expected
range of 1.0--1.2$\times 10^{22}$\,cm$^{-2}$ \citep{Keohane96}.

Even with this degeneracy some preliminary conclusions
are possible: i) the integrated Fe/O abundance ratio is
greater than 0.2 solar and ii) there is substantial He (and/or C, N)
compared to O$+$Fe, with $M$(O+Fe)$ < 0.01\,M$(He).
We can conclude that Cas\,A is dominated by O burning
\textit{products} (Si, S, etc.) rather than by oxygen \textit{per se}.

As a caveat, note that given the large region extracted and the unavoidable 
inclusion of a large radial range, which includes the blastwave,
it is possible that a finite contribution to the measured continuum
comes from synchrotron radiation~\citep{Hughes00}.

%

\begin{table*}[!t]\centering
  \newcommand{\DS}{\hspace{6\tabcolsep}} 
  \setlength{\tabnotewidth}{\textwidth}
  \setlength{\tabcolsep}{1.0\tabcolsep}
  \tablecols{12}
  \caption{Model\tabnotemark{a} Fits to
           the Oxygen and Fe-L  Emission from the N\,\&\,NW Region}\label{tab:fits}
  \begin{tabular}{cc l ccccc l ccc}
    \toprule
  \multicolumn{2}{c}{Fixed Values} &   & \multicolumn{4}{c}{Fit Parameters}  &  
     &   & \multicolumn{3}{c}{Derived Values\tabnotemark{c}}\\
  \cmidrule(l){1-2}\cmidrule(r){4-7}\cmidrule(l){10-12}
kT  &    $\tau$   & & 
  $N_{\mathrm H}$  &  Fe/O  &  O/He  &   $X_{\mathrm{norm}}$\tabnotemark{b}   &  $\chi^2$ & &  
     $n_e$  &  $M_\mathrm{He}$  &  $ \tau / n_e $  \\
(keV)   &   ($10^{11}$\,s\,cm$^{-3}$)  &  &
  ($10^{22}$\,cm$^{-2}$)  &     &      &  ($10^{-3}$)   &     &   &   
     (cm$^{-3}$)     &   ($M_\mathrm{solar}$)     &  (years)  \\
    \midrule
0.7  &     0.16  &  &  1.05  &  0.38  &  0.017  &  18.1  &  1.18  & &  
  73.  & 11.2  & 6.9 \\
0.7  &     0.20  &  &  1.0  &  0.46  &  0.016  &  14.5  &  1.11  & &  
  65.  & 10.0  & 9.7 \\
0.7  &    0.30  &  &  0.9  &  0.68  &  0.013  &   9.9  &  1.08  & &  
  54.  & 8.3  &  17. \\
0.7  &     0.60  &  &  0.8  &  0.82  &  0.013  &   7.0  &  1.10  & &  
  45.  & 7.0  &  42. \\
0.7  &     1.30  &  &  0.7  &  0.53  &  0.031  &   4.8  &  1.08  & &  
  38.  & 5.8  &  109. \\
0.7  &     3.00  &  &  0.7  &  0.15  &  0.154  &   4.3  &  1.05  & &  
  36.  & 5.4  &  266. \\
    \midrule
0.9  &     0.20  &  &  0.9  &  0.69  &  0.012  &   8.6  &  1.08  & &  
  50.  & 7.7  &  12.5 \\
0.9  &     0.60  &  &  0.6  &  1.98  &  0.014  &   2.2  &  1.04  & &  
  25.  & 3.9  &  74. \\
0.9  &     3.00  &  &  0.4  &  0.64  &  0.234  &   0.78  &  0.93  & &  
  15.  & 2.3  &  623. \\
    \midrule
1.2  &     0.16  &  &  0.8  &  1.03  &  0.010  &   5.0  &  1.07  & &  
  38.  & 5.9  &  13.1 \\
1.2  &     0.60  &  &  0.4  &  4.87  &  0.016  &   0.74  &  0.94  & &  
  15.  & 2.3  &  128. \\
1.2  &     3.00  &  &  0.6  &  0.69  &  0.290  &   2.9  &  1.04  & &  
  29.  & 4.4  &  323. \\
    \bottomrule
    \tabnotetext{a}{The model is an XSPEC {\texttt{wabs*vnei}} with
$kT$ and $\tau$ fixed as tabulated.  
The elemental abundances are: H=1, He=1000, O and Fe are free, and 
all others are set to 0.0.}
    \tabnotetext{b}{$X_{\mathrm{norm}}$ is the usual XSPEC model ``norm'' value;
see Paper\,I Appendix~B for details.}
    \tabnotetext{c}{These physical parameters are based on an
emitting volume equal to the cross-dispersion width (128\arcsec)
times one quadrant of the annulus from 100\arcsec to 130\arcsec:
$V = 0.25 * 128 * \pi(130^2-100^2)~~\approx$~9.3$\times 10^{55}$\,cm$^{3}$.
The equations of Paper\,I Appendix~B are
used with a distance to Cas\,A of 3.4\,kpc.}
  \end{tabular}
\end{table*}


\section{Future work}
\label{sec:future}

The current HETG data has produce a small but
tantalizing set of measurements on bright knots in Cas\,A
and given some information on the oxygen present in the
remnant.  Some additional work in these directions is
listed here:

\begin{itemize}

\item{It should be possible to measure the Doppler shifts
of additional knots in this data set as well as study individual
bright oxygen features.}

\item{We can compare the (sparse) 3D X-ray structure here with the optical/IR
``ring'' structures~(\citet{Reed95} and unpublished additional work.)}

\item{We have not yet examined the 
dispersed Fe-K emission in this data to see if it
contains useful information; if the Fe is indeed in fuzzy bubbles
then our slitless dispersive data will have reduced spectral
resolving power.}

\item{The non-dispersed \textit{Chandra} 1\,Ms data set is a unique resource
and these grating results can be 
used to confirm/calibrate the CCD-measured Doppler shifts.}

\item{It will be useful to create a ``Cas\,A knot catalog''
to allow comparisons across analyses, wavelengths, and epochs.}

\item{Although \textit{Suzaku} lacks the spatial resolution to isolate individual
knots, it does have better energy resolution in the oxygen 
band and may help constrain global Cas\,A emission models.}

\item{X-ray features in Cas\,A are changing in time~\citep{Patnaude06}
and a future HETG re-observation
($\Delta t > 6$~years) could give useful information on knot evolution.}

\end{itemize}

In the modeling domain, the analyses presented here suggest a larger goal
of building an approximate 3D model of the Cas\,A remnant (at a given epoch)
from a set of physically-consistent emission components.
Work with this flavor is already underway in astrophysics,
e.g., in the modeling of planetary nebulae \citep{Morisset05,Steffen06},
and will be accelerated by making more use of the 3D capabilities of
current computer hardware and software.

Finally, in the more distant hardware future,
one might imagine a ``soft X-ray spectrometer''
with a long-slit configuration operating in this low-energy
range\footnote{A minimum range of 14.5\,\AA~to~22.5\,\AA\ 
would include important Fe-L and O lines.  Extending shortward
to include \ion{Ne}{10} (12\,\AA) and/or longward to include
C and N lines (26-30-35-42\,\AA), would of course be desirable.}
to better isolate and measure the emission from complex extended sources
like Cas\,A.

In all this we are guided by
our current theories, but the data have the last word
and there may well be
some ``non-spherical-elephant'' surprises along the way.

\vspace{0.15in}

We thank Claude R. Canizares for providing GTO time for this observation and
Vikram Dwarkadas for providing 2D R-T data.
Thanks to Martin Laming and Rob Fesen for useful conversations.
Support for this work was provided by NASA/USA
through the Smithsonian Astrophysical Observatory (SAO)
contract SV3-73016 to MIT for Support of the \textit{Chandra} X-Ray Center,
which is operated by the SAO for and
on behalf of NASA under contract NAS8-03060.


\begin{thebibliography}

\bibitem[Bleeker et al.(2001)]{Bleeker01} Bleeker, J.~A.~M., 
Willingale, R., van der Heyden, K., Dennerl, K., Kaastra, J.~S., 
Aschenbach, B., \& Vink, J.\ 2001, \aap, 365, L225

\bibitem[Canizares et al.(2005)]{hetg05} Canizares, C.~R., et 
al.\ 2005, \pasp, 117, 1144
 
\bibitem[Chevalier et al.(1992)]{Chevalier92} Chevalier, R.~A., 
Blondin, J.~M., \& Emmering, R.~T.\ 1992, \apj, 392, 118 

\bibitem[Chevalier \& Oishi(2003)]{ChevOishi} Chevalier, R.~A., 
\& Oishi, J.\ 2003, \apjl, 593, L23

\bibitem[DeLaney et al.(2004)]{DeLaney04} DeLaney, T., Rudnick, 
L., Fesen, R.~A., Jones, T.~W., Petre, R., \& Morse, J.~A.\ 2004, \apj, 
613, 343

\bibitem[Dewey(2002)]{Dewey02} Dewey, D.\ 2002, High Resolution 
X-ray Spectroscopy with XMM-Newton and Chandra, 14D (ADS)

\bibitem[Dwarkadas(2000)]{Dwarkadas00} Dwarkadas, V.~V.\ 2000, 
\apj, 541, 418


\bibitem[Houck(2002)]{Houck02} Houck, J.~C.\ 2002, High 
Resolution X-ray Spectroscopy with XMM-Newton and Chandra, 17H (ADS)

\bibitem[Hughes et al.(2000)]{Hughes00} Hughes, J.~P., Rakowski, 
C.~E., Burrows, D.~N., \& Slane, P.~O.\ 2000, \apjl, 528, L109

\bibitem[Keohane et al.(1996)]{Keohane96} Keohane, J.~W., 
Rudnick, L., \& Anderson, M.~C.\ 1996, \apj, 466, 309


\bibitem[Laming \& Hwang(2003)]{Laming03} Laming, J.~M., \& 
Hwang, U.\ 2003, \apj, 597, 347

\bibitem[Lazendic et al.(2006)]{Lazendic06} Lazendic, J.~S., 
Dewey, D., Schulz, N.~S., \& Canizares, C.~R.\ 2006, \apj, 651, 250 
(Paper\,I)

\bibitem[Markert et al.(1983)]{Markert83} Markert, T.~H., Clark, 
G.~W., Winkler, P.~F., \& Canizares, C.~R.\ 1983, \apj, 268, 134

\bibitem[Morisset et al.(2005)]{Morisset05} Morisset, C., 
Stasi{\'n}ska, G., \& Pe{\~n}a, M.\ 2005, \mnras, 360, 499

\bibitem[Patnaude \& Fesen(2006)]{Patnaude06} Patnaude, D.~J., \&
Fesen, R.~A.\ 2006, \aj, accepted; astro-ph/0609412

\bibitem[P{\'e}rez-Rend{\'o}n et al.(2002)]{PerezRendon02} 
P{\'e}rez-Rend{\'o}n, B., Garc{\'{\i}}a-Segura, G., \& Langer, N.\ 2002, 
RevMexAA, 12, 94

\bibitem[Reed et al.(1995)]{Reed95} Reed, J.~E., Hester, 
J.~J., Fabian, A.~C., \& Winkler, P.~F.\ 1995, \apj, 440, 706

\bibitem[Steffen \& L{\'o}pez(2006)]{Steffen06} Steffen, W., \& 
L{\'o}pez, J.~A.\ 2006, RevMexAA CS, 26, 30


\end{thebibliography}
\end{document}